\documentclass{ws-procs11x85}
\usepackage{balance} 

\begin{document}

\twocolumn[
\title{LS I+61 303: microquasar or not microquasar?}

\author{G. E. Romero$^*$ and M. Orellana}

\address{Instituto Argentino de Radioastronom\'{\i}a, C.C. 5,\\
(1894) Villa Elisa, Buenos Aires, Argentina \\
$^*$E-mail: romero@fcaglp.unlp.edu.ar\\
www.iar.unlp.edu.ar/garra}

\author{A. T. Okazaki}

\address{Faculty of Engineering, Hokkai-Gakuen University,\\
Toyohira-ku, Sapporo 062-8605, Japan\\
E-mail: okazaki@elsa.hokkai-s-u.ac.jp}

\author{S. P. Owocki}

\address{Bartol Research Institute, University of Delaware,\\
Newark, DE 19716, USA\\
E-mail: owocki@bartol.udel.edu}

\begin{abstract}
LS I +61 303 is a puzzling object detected from radio up to
high-energy gamma-rays.  Variability has recently been observed in its
high-energy emission.  The object is a binary system, with a compact
object and a Be star as primary.  The nature of the secondary and the
origin of the gamma-ray emission are not clearly established at
present.  Recent VLBA radio data have been used to claim that the
system is a Be/neutron star colliding wind binary, instead of a
microquasar.  We review the main views on the nature of LS I +61
303 and present results of 3D SPH simulations that can shed
some light on the nature of the system.  Our results support an
accretion powered source, compatible with a microquasar
interpretation.
\end{abstract}
\vskip12pt
]

\bodymatter


\section{Introduction}\label{int}

The radio-emitting X-ray binary LS I +61 303 was discovered by Gregory
\& Taylor (1978).  It is formed by a primary B0-B0.5Ve star with a
dense equatorial disk (Paredes \& Figueras 1986) and a compact object
of unknown nature (Casares et al.  2005).  The orbital period is
29.4960 days (Gregory 2002).  The eccentricity is estimated to be
high, ranging from $\sim0.55$ (Grunsdstrom et al.  2007) to
$0.72\pm15$ (Casares et al.  2005).  The distance to the system is
$\sim 2$ kpc (e.g. Frail \& Hjellming 1991).

At X-ray energies the source presents a behavior different from other
X-ray binaries: the overall level of emission is rather low ($\sim
10^{33}-10^{34}$ erg s$^{-1}$) and no outbursts are observed.  The
whole spectral energy distribution (SED) has been recently compiled by
Sidoli et al.  (2006).  No pulsation is observed from the compact
object in any band.  Thermal X-ray features are absent as well, with
the spectrum following a power-law.

Recently, LS I +61 303 was detected by the MAGIC Cherenkov telescope
at energies $E>200$ GeV (Albert et al.  2006).  The emission is highly
variable and modulated by the orbital period.  The maximum flux is
observed well after the periastron passage, at phase $\sim0.6$ (the
periastron is at phase 0.23).  The gamma-ray emission and variability
have been confirmed by the VERITAS array (Maier et al.  2007).  The source
was also detected by EGRET at $E>100$ MeV (Kniffen et al.  1997), with
a luminosity about $\sim 10^{35}$ erg s$^{-1}$.

High-resolution studies of the radio morphology have shown the
existence of jet-like features that change very rapidly, on timescales
of days (Massi et al.  2001, 2004).  These observations were first
considered as supportive of a microquasar scenario (Massi 2004), but
recently Dhawan et al.  (2006) claimed that the changing radio
morphology revealed by the VLBA along the orbit is evidence for a
Be/pulsar system, powered by wind collisions as it is the case in PSR
1259-63, a well-known system that also produces high-energy emission
(Aharonian et al.  2005).  In this paper we discuss the nature of
this peculiar source, in light of the muliwavelength observations
and through detailed 3D SPH simulations of both the colliding-wind and
accretion-jet scenarios.

\section{Colliding winds and cometary tails}

The scenario based on colliding winds for LS~I~+61~303 was originally
proposed by Maraschi \& Treves (1981), and recently revisited by Dubus
(2006).  Dhawan et al.  (2006) claim that a jet-like feature observed
during the periastron passage could be a `cometary tail' produced by
the stellar wind around the pulsar.  The stellar wind, in this
interpretation, would confine the pulsar wind producing a jet-like
radio morphology that would point opposite to the star.  Relativistic
electrons, locally accelerated at the bow shock, would cool via
synchrotron radiation and inverse Compton interactions in this
scenario.  The absence of thermal features in the X-ray spectrum is
used to argue against the existence of accretion in the system.  But the
wind collision picture has several problems that require further
investigation.  For example, the `cometary tail' points in random
directions at phases other than the periastron, even in the direction
of the Be star at certain phases.  A spin-down luminosity of the
pulsar of ($\sim 10^{36}$ erg s$^{-1}$) would require unrealistic
efficiencies for the generation of gamma-rays in the GeV band.  Larger
pulsar luminosities would produce a wind that should overcome the
stellar wind and would require a too-young pulsar to sustain an
extremely powerful relativistic wind.  Finally, in this colliding-wind 
scenario, it is difficult to explain the multiwavelength light
curve, something that can be done in a scenario with accretion and
associated ejection of unstable outflows (e.g. Bosch-Ramon et al.
2006).

\begin{figure}[]
\center
\centerline{\psfig{figure=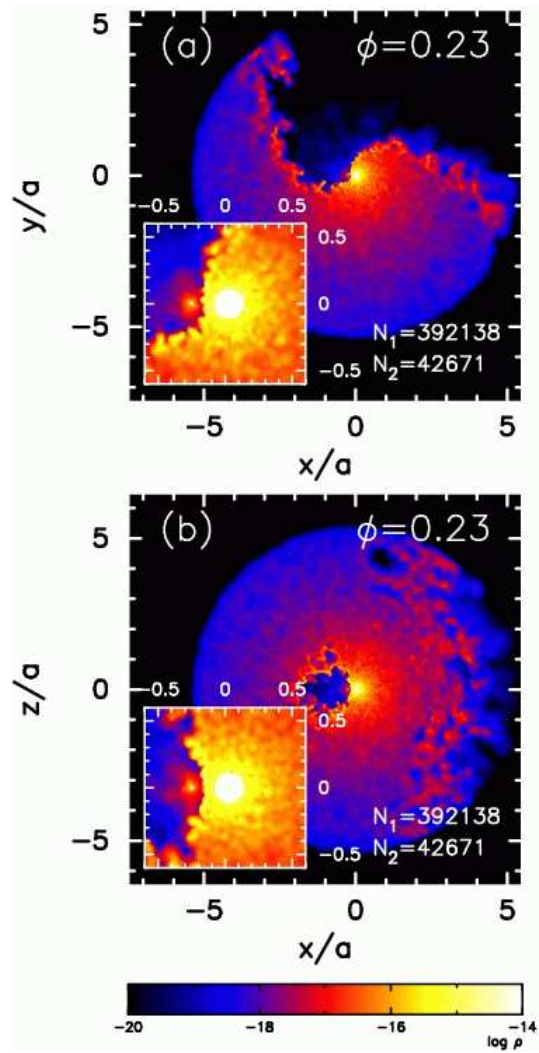,width=7truecm}}
\caption{
Colliding wind snapshots at periastron from the 3D SPH sumulations of
LS I+61 303.  The bright spot is the Be star whereas the small point
is the pulsar.  The orbit is in the $xy$ plane.  Lengths are
in units of the semi-major axis $a$ of the orbit.  \label{f1}}
\end{figure}

A helpful way to check the colliding wind scenario is to
perform 3D time-dependent numerical simulations of the system under
the assumptions made by the proponents of the Be/pulsar hypothesis.
We have done such simulations (described in detail in Romero et al.
2007) using a well-tested SPH code (e.g. Okazaki et al.  2002).  We
assumed a rapid Be wind with a velocity of 1000 km s$^{-1}$ and a
mass loss rate of $10^{-8}$ $M_{\odot}$ yr$^{-1}$.  The ratio of
momentum fluxes is then $\eta=0.53 \dot{E}_{\rm PSR}/10^{36}$ erg
s$^{-1}$, where $\dot{E}_{\rm PSR}$ is the pulsar spin-down
luminosity.  The pulsar wind was assumed to have a velocity of
$10^{4}$ km s$^{-1}$ in order to avoid the complexities of a
relativistic wind, but the mass-loss rate was adjusted as to provide
the same momentum as a relativistic flow.  Our results, shown in Fig.
\ref{f1}, are in agreement with 2D non-dynamical but relativistic
simulations performed by Bogovalov et al.  (2007).  There is no evidence
of any cometary tail for a pulsar power of $\sim 10^{36}$ erg s$^{-1}$.  
Moreover, the Be stellar wind cannot confine
the pulsar wind. For a lower power, the energetics of
the gamma-ray emission cannot be accounted for.  For a higher power,
the pulsar should be too young, and the shock front would wrap around
the star, not the pulsar.

\section{What is the accretion regime of LS~I+61~303?}

\begin{figure}[]
\center
\centerline{\psfig{figure=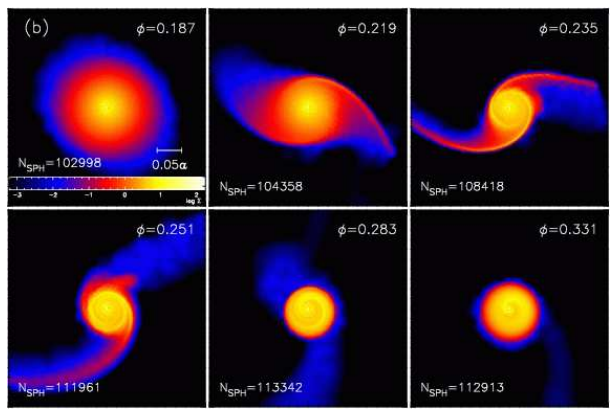,width=8truecm}}
\caption{
Evolution of the accretion disk close to the periastron passage
according to the simulations.  The density wave excited by the mass
transfer is clearly seen.\label{f2}}
\end{figure}

\begin{figure}[]
\center
\centerline{\psfig{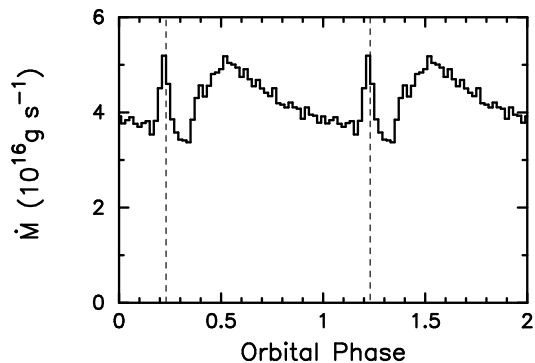}}
\caption{
Accretion rate obtained through the simulations.  Notice that the
accretion is sub-Eddington, in a regime where ADAF, sub-luminous,
solutions exist.\label{f3}}
\end{figure}

We have performed full simulations of the mass transfer from the Be
star and its circumstellar disk to the compact object under the
assumption that the latter is a 2.5 $M_{\odot}$ black hole (see
details in Romero et al.  2007).  The bulk of the mass flux onto the
compact object occurs around periastron passage.  The black hole
gravitational pull tidally deforms the circumstellar disk and particles are
captured and accreted, forming an accretion disk in which a strong
density wave is excited (see Fig.  \ref{f2}).  When the wave arrives
near the central hole, around phase 0.5, there is a broad peak in the 
accretion rate, as shown in Fig.  \ref{f3}.  This peak is correlated with 
the maximum of the gamma-ray emission in accretion/ejection models (e.g.
Orellana \& Romero 2007), through the standard jet/disk symbiosis
assumption.  The accretion rate resulting from the simulations is
significantly lower (by more than two orders of magnitude) than what
is obtained through simple calculations based on Bondi-Hoyle
accretion into a non-perturbed circumstellar disk ( Mart\'{\i} \&
Paredes 1995, Gregory \& Neish 2002).  It is very interesting to
note that for the accretion rates obtained from the SPH simulations,
the accretion regime would be advection-dominated (Narayan et al.
1999).  This has important consequences, since the source should then
be underluminous in X-rays, and the nonthermal emission could easily
cover any thermal feature from the accretion disk.

The narrow accretion peak at the periastron might be, as explained by
Romero et al.  (2007), an artifact of the spatial resolution of the
simulations.  In any case, when opacity effects to gamma-ray
propagation in the anisotropic radiation field of the star and the
truncated circumstellar disk are taken into account, the high-energy
emission during the periastron passage must be strongly attenuated.

\section{Where are the Be/black hole binaries?}

\begin{figure}[]
\center
\centerline{\psfig{figure=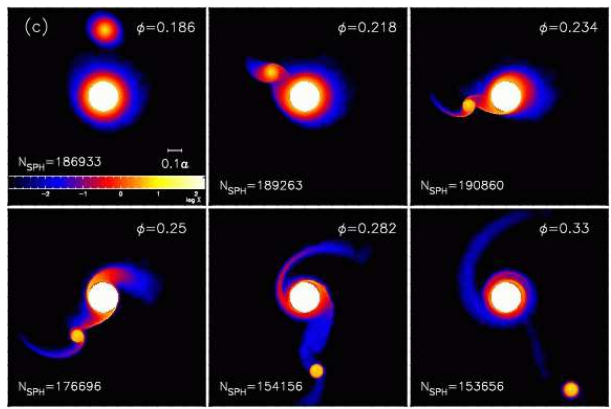,width=8truecm}}
\caption{
Tidal interactions between the compact object and the circumstellar
disk around the periastron passage.  \label{f4}}
\end{figure}

Known Be/X-ray binaries harbor pulsars and have long periods.
However, Be/black hole binaries should exist in the Galaxy according
to population synthesis studies (e.g. Podsiadlowski et al.  2003).
They are expected to be narrow systems, with orbital periods of less
than 30 days (Zhang et al.  2004).  The tidal effects that truncate
the Be disk described by our simulations make these systems
underluminous in X-rays, and hence far more difficult to detect than
longer period Be/pulsar binaries, where large X-ray outbursts use to
occur.  In Fig.  \ref{f4} we show the effects of the tidal truncation
of the Be disk.  In other systems with a pulsar and a long period
(e.g. PSR 1259--63), the disk has  enough time to expand to large
distances after the interaction with the compact object, which then
moves {\sl through} the disk in the next approach.  We emphasize
then that LS I +61 303 is a very different system from PSR 1259-63.
We suggest that LS I +61 303 could be the first Be/black hole binary
detected in the Galaxy.

\section{Conclusions: LS I +61 303, a peculiar microquasar}

The conclusion of our work, contrary to some recent claims found in
the literature, is that LS I +61 303 can be classified as a
microquasar because it is powered by accretion and presents radio
emission produced by what seems to be an unstable jet.  It is a quite
different system from PSR 1259-63.  The instability of the jet might
be the result of the interaction of the outflow with the wind (Perucho
\& Bosch-Ramon 2007) or the effects of perturbations in the accretion
disk and the attached magnetic fields (Romero et al.  2007).  The
overall picture, hence, should be similar to that discussed by Massi
(2004).  The mechanism behind the gamma-ray production might be
hadronic (Romero et al.  2003, 2005; Orellana \& Romero 2007),
leptonic (Bosch-Ramon et al.  2006, Bednarek 2006), or a mixture.
Neutrino observations with ICECUBE could settle this issue in the near
future.

\subsection*{Acknowledgements}

We thank V. Bosch-Ramon and D. Khangulyan for isightful discussions. 
G.E.R. \& M.O. are supported by the Argentine Agencies CONICET 
(GRANT PIP 5375) and ANPCyT (GRANT PICT 03-13291 BID 1728/OC-AR). 
The former is also supported by the Ministerio de Eduaci\'on y Ciencia 
(Spain) under grant AYA2007-68034-C03-01, FEDER funds.
A.T.O. thanks the Bartol Research Institute, University
of Delaware, USA for the warm hospitality during his sabbatical visit.
He also acknowledges Japan Society for the Promotion of Science for
the financial support via Grant-in-Aid for Scientific Research
(16540218).
SPH simulations were performed on HITACHI SR11000 at Hokkaido
University Information Initiative Center and SGI Altix~3700 at
Yukawa Institute of Theoretical Physics, Kyoto University.
S.P.O. acknowledges partial support of NSF grant 0507581. 

\balance



\begin{thebibliography}{99}
\bibitem{}Aharonian, F. A., et al. (HESS Coll.), 2005, A\&A, 442, 1
\bibitem{}Albert, J. et al. (MAGIC coll.) 2006, Science, 312, 1771
\bibitem{}Bednarek, W., 2006, MNRAS, 371, 1737 
\bibitem Bogovalov, S., et al., 2007, [arXiv:0710.1961] 
\bibitem{}Bosch-Ramon, V., Paredes, J.M., Romero, G.E., \& Rib\'o, M., 2006, A\&A, 459, L25
\bibitem{}Casares, J., et al., 2005, MNRAS, 360, 1105
\bibitem{}Dhawan, V., Mioduszewski, A., \& Rupen, M., 2006, in {Proc.
of the VI Microquasar Workshop}, Como-2006
\bibitem{}Dubus, W., 2006, A\&A, 456, 801
\bibitem{}Frail, D. A. \& Hjellming, R. M., 1991, AJ, 101, 2126
\bibitem{}Gregory, P.C, \& Taylor, A.R., 1978, Nature, 272, 704 
\bibitem{}Gregory, P.C., \& Neish, C., 2002, ApJ, 580, 1133
\bibitem{}Gregory, P.C., 2002, ApJ, 575, 427 
\bibitem{}Grundstrom, E. D., et al. 2007, ApJ, 656, 437   
\bibitem{}Kniffen, D.A., et al., 1997, ApJ, 486, 126 
\bibitem{}Maier, G., et al. (VERITAS Coll.), 30th ICRC, 2007. 
\bibitem{}Maraschi, L. \& Treves, A., 1981, MNRAS 194, 1
\bibitem{}Mart\'{\i}, J., \& Paredes, J.M., 1995, A\&A, 298, 151 
\bibitem{}Massi, M., et al., 2001, A\&A, 376, 217  
\bibitem{}Massi, M., et al., 2004, A\&A 414, L1 
\bibitem{}Massi, M., 2004, A\&A 422, 26 
\bibitem{}Narayan, R., Mahadevan, R., \& Quataert, E., 1998, in: M.A. Abramowickz, 
G. Bj\"ornsson, \& J.E. Pringle (eds), Theory of Black Hole Accretion Disks, 
Cambridge University Press, Cambridge, p. 148. 
\bibitem{}Okazaki, A.T., Bate, M.R., Ogilvie, G.I \& Pringle J.E., 2002, MNRAS, 337, 967
\bibitem{}Orellana, M., \& Romero, G.E., 2007, Ap\&SS, 309, 333
\bibitem{}Paredes, J.M., \& Figueras, F., 1986, A\&A, 154, L30
\bibitem{}Perucho, M, \& Bosch-Ramon, V., 2007, [arXiv:0710.3556]
\bibitem{}Podsiadlowski, P., Rappaport, S. \& Han, Z., 2003, MNRAS, 341, 385
\bibitem{}Romero, G.E., et al., 2003, A\&A, 410, L1
\bibitem{}Romero, G.E., Christiansen, H.R., \& Orellana, M., 2005, ApJ, 632, 1093
\bibitem{}Romero, G.E., et al., 2007, A\&A, 474, 15
\bibitem{}Sidoli, L. et al., 2006, A\&A, 459, 901
\bibitem{}Zhang, F., Li, X.D., \& Wang, Z. -R., 2004, ApJ, 603, 663  

\end{thebibliography}
\end{document}